 \documentclass[smallabstract,smallcaptions]{dccpaper}
\pdfoutput=1
\usepackage{epsfig}
\usepackage{amsmath}
\usepackage{amssymb}
\usepackage{color}
\usepackage{url}

\newlength{\figurewidth}
\newlength{\smallfigurewidth}

\newcommand{\etal}{\textit{et al.}}
\usepackage{subfigure}
\usepackage{graphicx}
\usepackage{caption}
\usepackage{multirow}
\usepackage{multicol}
\usepackage{booktabs}
\usepackage{hyperref}
\begin{document}

\title
{\large
\textbf{An Efficient QP Variable Convolutional Neural Network Based In-loop Filter for Intra Coding}
}

\author{%
Zhijie Huang, Xiaopeng Guo, Mingyu Shang, Jie Gao and Jun Sun$^{\ast}$\thanks{$^{\ast}$Corresponding author}\\[0.5em]
{\small\begin{minipage}{\linewidth}\begin{center}
\begin{tabular}{ccc}
% $^{\ast}$
Wangxuan Institute of Computer Technology \\
Peking University \\
Beijing, 100871, China \\
\url{{zhijiehuang, jsun}@pku.edu.cn}
\end{tabular}
\end{center}\end{minipage}}
}

\maketitle
\thispagestyle{empty}

\begin{abstract}
In this paper, a novel QP variable convolutional neural network based in-loop filter is proposed for VVC intra coding. To avoid training and deploying multiple networks, we develop an efficient QP attention module (QPAM) which can capture compression noise levels for different QPs and emphasize meaningful features along channel dimension. Then we embed QPAM into the residual block, and based on it, we design a network architecture that is equipped with controllability for different QPs. To make the proposed model focus more on examples that have more compression artifacts or is hard to restore, a focal mean square error (MSE) loss function is employed to fine tune the network. Experimental results show that our approach achieves 4.03\% BD-Rate saving on average for all intra configuration, which is even better than QP-separate CNN models while having less model parameters.
\end{abstract}

\Section{1. Introduction}

In-loop filtering is an essential module in video coding, which can not only improve the quality of current frames directly by reducing the compression artifacts but also provide high-quality reference frames for succeeding pictures. In the latest video coding standard Versatile Video Coding (VVC) \cite{VVC}, four in-loop filtering steps, namely a luma mapping with chroma scaling (LMCS) process \cite{LMCS}, followed by a deblocking filter (DBF) \cite{DBF}, an SAO filter \cite{SAO} and an adaptive loop filter (ALF) \cite{ALF} are applied to the reconstructed samples. The DBF and SAO are similar to that of the HEVC \cite{HEVC} standard, whereas LMCS and ALF are newly adopted in VVC.

Besides the built-in in-loop filters in video coding, various convolutional neural network (CNN) based in-loop filters have been proposed in recent years. In \cite{RRCNN}, a very deep recursive residual CNN (RRCNN) was developed to recover the reconstructed intra frames. Zhang {\etal}  \cite{RHCNN} introduced a deep residual highway CNN (RHCNN) based in-loop filtering in HEVC. Wang {\etal} \cite{DRNLF} designed a dense residual CNN based in-loop filter (DRNLF) for VVC. Typically, since the compression noise levels are distinct for videos compressed with different quantization parameters (QPs), we need to train many CNN models for different QPs. To address this issue, Zhang {\etal} \cite{RHCNN} merged the QPs into several bands, and trained the optimal models for each band, but they still had to train and deploy several networks. Song {\etal} \cite{CNNLF} combined QPs as an input and fed them into the CNN training stage by simply padding the scalar QPs into a matrix with the same size of input frames or patches. However, these QP-combined models are inferior to QP-separate CNN models in terms of rate-distortion (RD) performance. And their flexibility and scalability are not strong.

In this paper, a novel QP variable convolutional neural network based in-loop filter is proposed for VVC intra frames. Specifically, Considering different compression noise levels for different QPs, a QP attention module (QPAM) is developed which assigns different weights to each channel of the input feature map according to the QP value. Compared with other methods, the proposed QPAM has wide applicability and stronger scalability, which can also be applied to adapt to different frame types. Then QPAM is embed into the residual block. Based on it, we design a network architecture that can not only fully utilize residual feature, but has the controllability for different QPs. To further make the proposed model pay more attention to examples that have more compression artifacts or is hard to restore, a focal mean square error (MSE) loss function is employed to fine tune the network. 
Experimental results verify the efficiency of the proposed QPAM and network architecture, which also outperforms other methods.

\Section{2. The Proposed QPALF Method}\label{method}
\begin{figure}[t]
\centering
\includegraphics[width=0.6\linewidth]{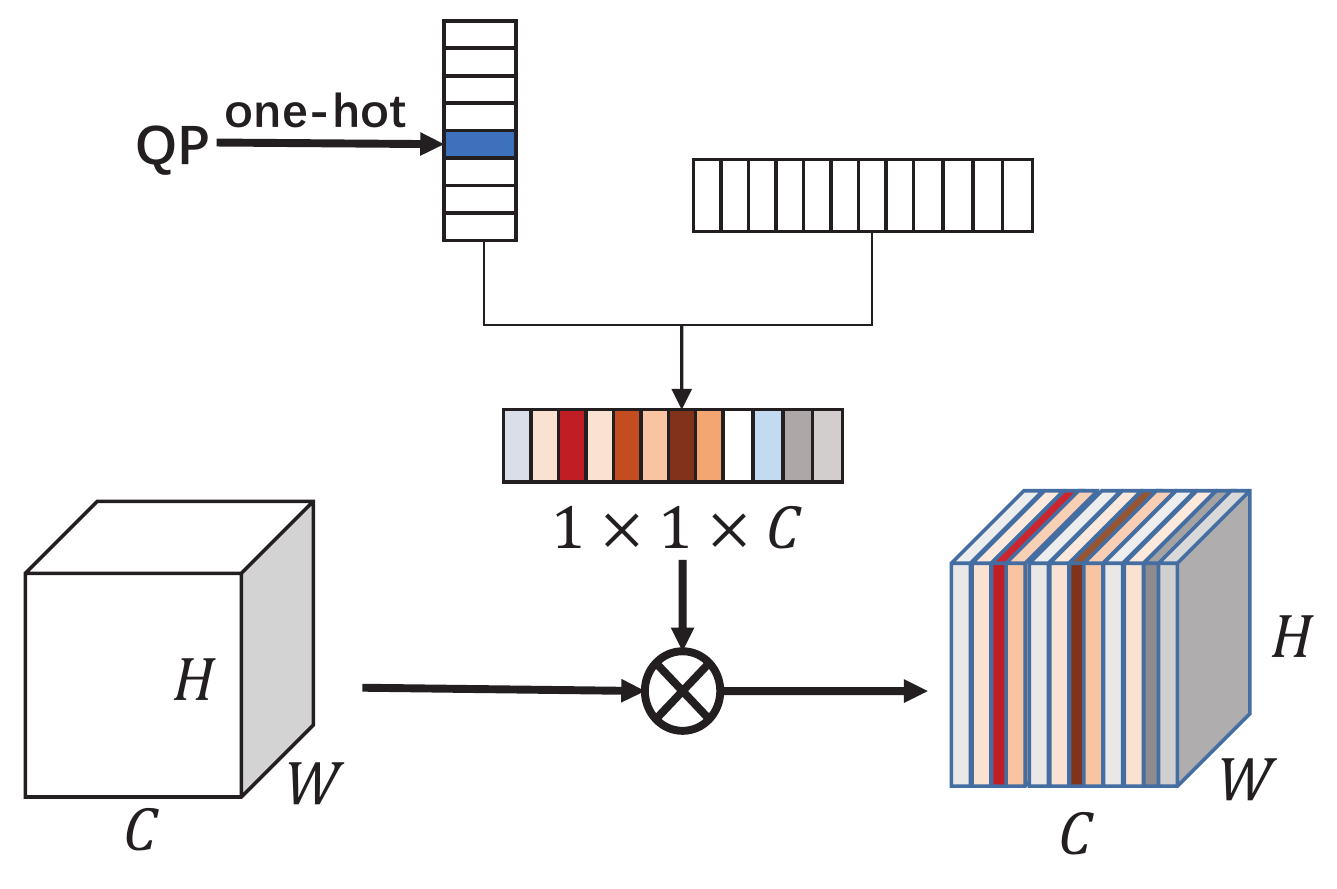}%
\caption{Overview of the QP attention module (QPAM). $\otimes$ denotes element-wise product.}
\label{QPA}
\end{figure}
\subsection*{QPAM}
Inspired by channel attention module in \cite{attention}, we propose a QPAM to avoid training and deploying multiple networks. Unlike channel attention module which extracts the attention map from the input feature map, our QP attention module is controlled by the QP value. The overview of the proposed QPAM is illustrated in Figure \ref{QPA}. Given a feature map $\mathbf{F} \in \mathbb{R}^{ H \times W  \times C}$ as input, QPAM sequentially infers a 1D QP attention map $\mathbf{M} \in \mathbb{R}^{ 1 \times 1  \times C}$. The attention process can be summarized as:
\begin{equation}
    \mathbf{F'} = \mathbf{M} \otimes \mathbf{F}  
\end{equation}
where $\otimes$ denotes element-wise multiplication. During multiplication, the attention values are broadcast along the channel dimension. $ \mathbf{F'}$ is the refined output.
The process of generating the QP attention map is as follows: Given a QP value $ q \in \Omega = [a, b]$, since $q$ is an integer, we first map $q$ to a vector $\mathbf{v}_{\Omega}(q) \in \mathbb{R}^{ m \times 1}$ by one-hot encoding. $m$ is the length of $|\Omega|$. The QP attention map $\mathbf{M}$ is calculated by:
\begin{equation}
\begin{array}{l}
    \mathbf{M'} = \sigma(U\mathbf{v}_{\Omega}(q)) \\
    \mathbf{M} = \text{reshape}(\mathbf{M'})
\end{array}
\end{equation}
where $U \in  \mathbb{R}^{ C  \times m}$ is a weight matrix and $\sigma(x)=log(1+e^x)$. From the process we can see that the QPAM assigns different weights to each channel of the input feature map according to the QP value, so that the module can capture compression noise levels among different QPs. Meanwhile, the module can also emphasize meaningful features along channel axes through this process.
Moreover, compared with other methods, the proposed QPAM has stronger scalability, which can also be easily applied to other discrete variables, e.g. frame types.
% \usepackage{graphicx}
% \usepackage{subfigure} %需要使用的宏包

% \begin{figure}
% \centering
% \subfigure[QPALF]{\includegraphics[width=\textwidth]{network-eps-converted-to.pdf}}

% \subfigure[Boot]{
%     \begin{minipage}[t]{\textwidth}
%     \includegraphics[width=1.5in]{img/RB-eps-converted-to.pdf}
%     \end{minipage}%

%     \begin{minipage}[t]{\textwidth}
%     \includegraphics[width=1.5in]{img/RFA-eps-converted-to.pdf}
%     \end{minipage}
% }

% \caption{Jackson Yee}
% \end{figure}

\subsection*{QPALF}
\begin{figure}[t]
\centering
\includegraphics[width=0.8\linewidth]{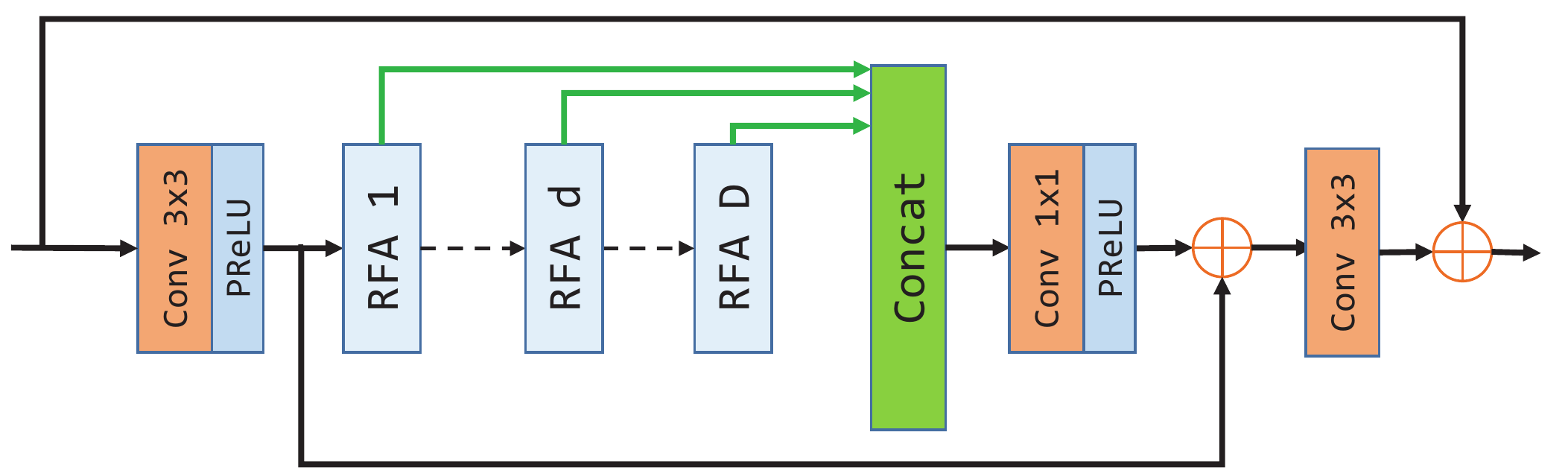}%
\caption{Overview of the architecture of our QPALF network.}
\label{network}
\end{figure}
\begin{figure}[htbp]
\centering
\subfigure{
\begin{minipage}[t]{0.5\linewidth}
\centering
\includegraphics[width=\linewidth]{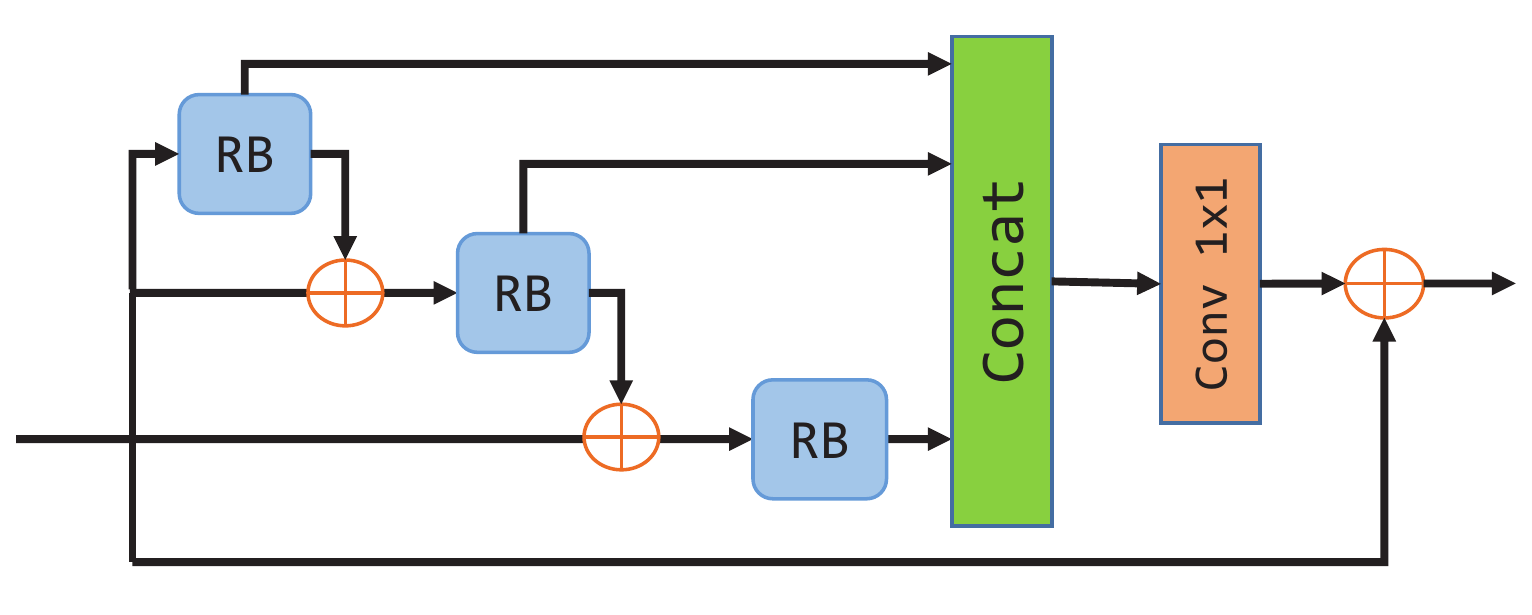}
% \caption*{Ours(\textbf{28.42, 0.8714})}
\label{RFA}
\end{minipage}
}%
\subfigure{
\begin{minipage}[t]{0.5\linewidth}
\centering
{\includegraphics[width=\linewidth]{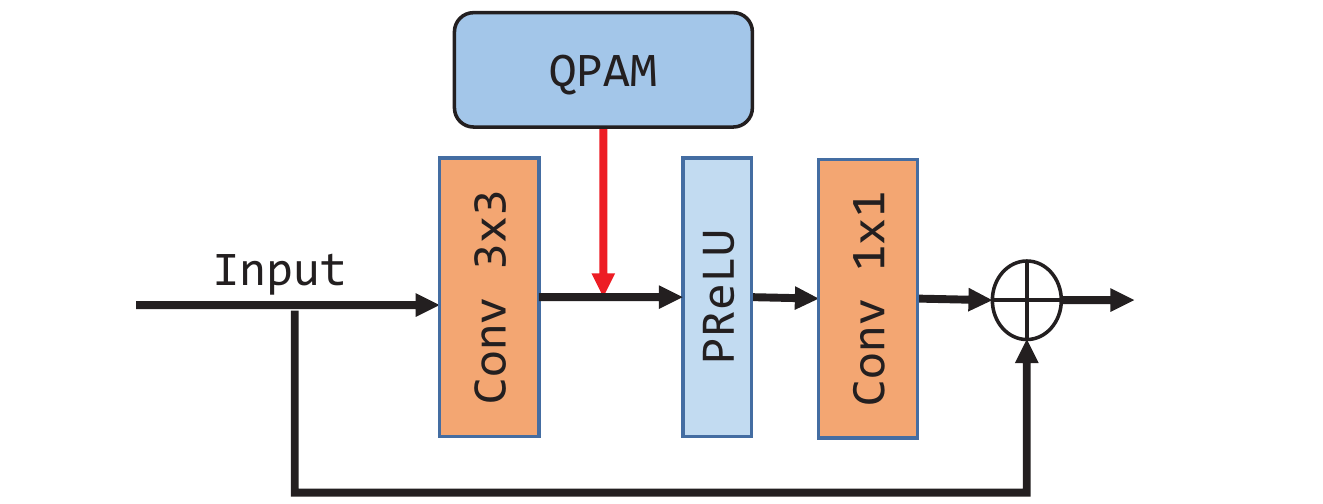}%
\label{RB}}
% \caption*{DRN(28.32, 0.8682)}
\end{minipage}
}%
\caption{Detail of Residual Block with QPAM and Residual Feature Aggregation Module. \textbf{Left}: (a) Residual Feature Aggregation Block. RB denotes Residual Block. \textbf{Right}: (b) Residual Block with QPAM.}
\end{figure}

\textbf{Architecture}. In figure \ref{network}, we present the overview of the architecture of our QPALF network. This is also one of the popular architecture used by many other methods \cite{RHCNN, RRCNN, DRNLF}, which usually consists of three parts: the head part, the backbone part and the reconstruction part. The head part is responsible for initial feature
extraction with only one convolutional layer followed by an activate function.
Given a compressed input $X$, we can get a shallow feature $F_0$ through this layer:
\begin{equation}
    F_0=\mathcal{F}(X)
\end{equation}
The backbone part is the key component of the network, which is also the most distinct part of various networks. Here it makes up of $D$ cascaded residual feature aggregation modules (RFA). The backbone part receives the feature $F_0$ as input and sends the extracted global feature $F$ to the reconstruction part, which can be formulated as:
\begin{equation}
    F_D= \mathcal{R}_D(F_{D-1})=\mathcal{R}_{D}\left(\mathcal{R}_{D-1}\left(\ldots\left(\mathcal{R}_{1}\left(F_{0}\right)\right) \ldots\right)\right)
\end{equation}
\begin{equation}
    F= F_0 + \mathcal{R}([F_1, \ldots, F_d, \ldots, F_D])
\end{equation}
where $\mathcal{R}_d$ denotes the d-th RFA module function. $F_{d-1}$ is the input feature of the d-th RFA module function and $F_{d}$ is the corresponding output. The output features of the $D$ RFAs are concatenated together. Then we utilize a long skip connection to extract a global feature $F$.
Finally, the global feature $F$ is transformed through the reconstruction part 
\begin{equation}
    \hat{Y} = X + \mathcal{H}(F)
\end{equation}
where $\hat{Y}$ is the output and $\mathcal{H}$ is the reconstruction function, which consists of only one convolutional layer. A global residual learning is usually used to ease the training difficulty in the reconstruction part.

Inspired by \cite{RFA}, we propose a RFA module to make a better use of the local residual features. Figure \ref{RFA} illustrates the detail of the RFA, which contains three residual blocks and one convolution layer. The input is extracted by three residual blocks at three different levels. Then the outputs from three residual blocks are concatenated and a $1 \times 1$ convolution is applied at the end of the RFA for channel dimension reduction. As depicted in Figure \ref{RB}, the detail of the residual block is the same as that in \cite{ResNet} except that we employ the proposed QPAM after the first convolution layer. 

Considering the complexity of the proposed model, we use 6 RFA modules, that is, $D=6$ here. All of $3 \times 3$ convolutional layers are followed by a parametric rectified linear unit (PReLU) \cite{PRelu} activation, and the filter number of each $3 \times 3$ layer convolutional is 64. The $1 \times 1$ convolutions are applied for channel dimension reduction.

\textbf{Dataset}. For the network training, we build a dataset using DIV2K \cite{NTIRE} which contains 800 high-resolution images. First, we convert these images to YUV 4:2:0 color format and encode them by VTM6.0 \cite{VTM} with all-intra (AI) configuration at four QPs, {22, 27, 32, 37}. The built-in in-loop filters are all enabled when compressed these images. Then the compressed images are divided into two non-overlapping sets of training (700 images), validation (100 images). To further expand the training dataset, we split the reconstructions to small patches of $64\times64$ with stride 16. And we remove the patches whose PSNR are more than 50.0 or less than 20.0. When training QP variable models, four training datasets are mixed in a random order.
\begin{figure}[t]
\centering
\includegraphics[width=0.5\linewidth]{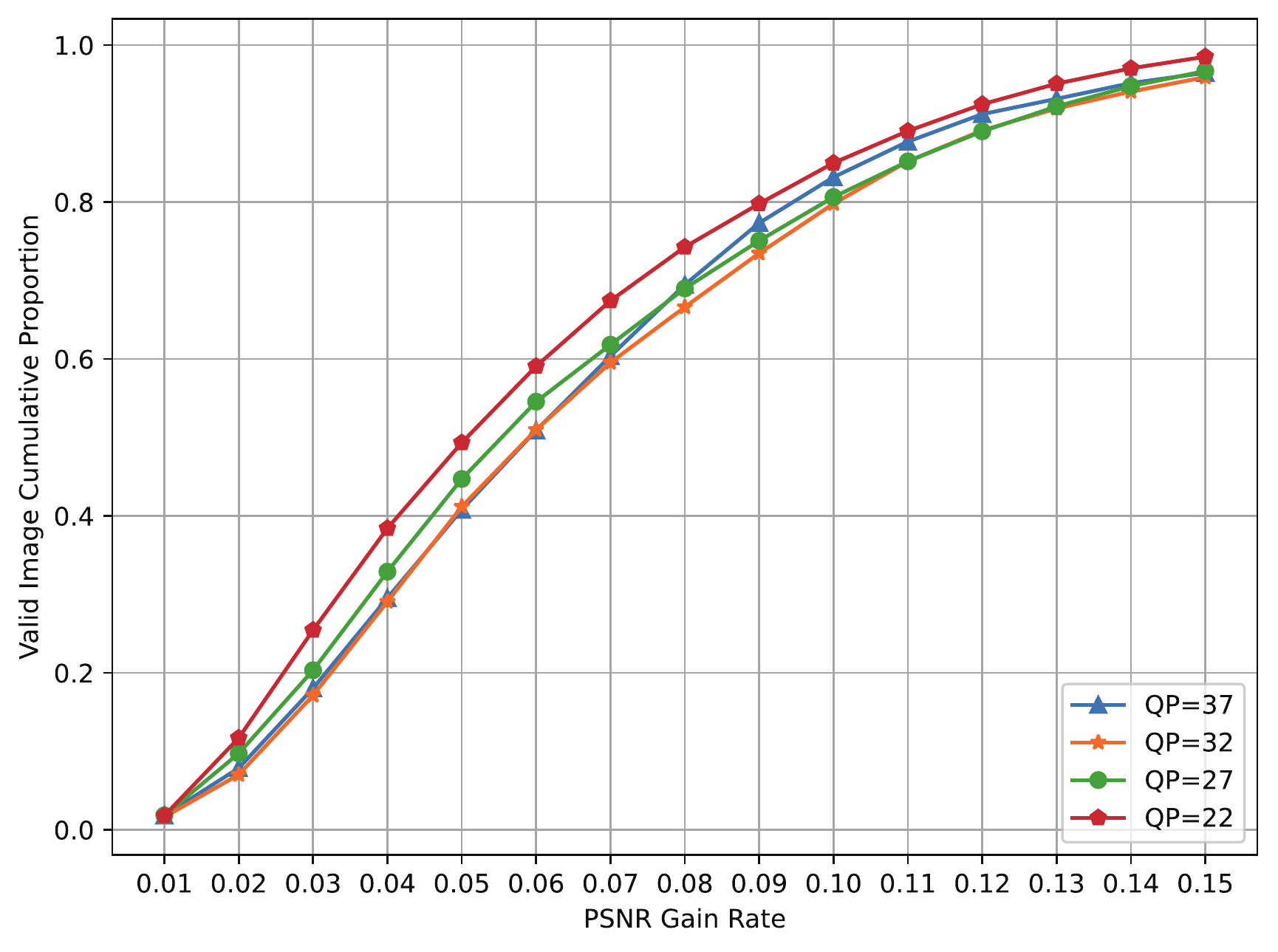}
\caption{Cumulative PSNR gain rate distribution of QPALF on valid image dataset.}
\label{culumative}
\end{figure}

\textbf{Loss Function}. Let $X$ be the input and $\theta$ be the set of network parameters to be optimized. Our goal is to learn an end-to-end function $F$ for generating a higher quality reconstruction $\hat{Y} = F(X;\theta)$ that is close to the ground truth $Y$. The loss function is the MSE between $\hat{Y}$ and $Y$:
\begin{equation}
L_{rec}=\frac{1}{N} \sum_{i=1}^{N}{\lVert \hat{Y}^{(i)} - Y^{(i)} \rVert ^2}
\end{equation}
where $N$ is the number of training samples in each batch. In order to train a more robust QP-combined network, we analyse the restoration ability of the network for different QPs. First we train a QPALF network using mixed dataset (the training detail will be presented in the follow). Then we plot the valid image cumulative proportion over the PSNR gain rate for different QPs. And the PSNR gain rate is defined as follows:
\begin{equation}
    R = 1 - \frac{L_{rec}}{L_{init}}
\end{equation}
where $L_{init}$ is the MSE between $X$ and $Y$. From Figure \ref{culumative}, we can find: 1) The network has lower PSNR gain rate overall on dataset with smaller QP, especially at QP=22. Obviously smaller QP means less compression artifacts, and we do not expect the network to pay much attention on data with less compression artifacts. 2) The PSNR gain rate of 80\% of the valid data is less than 10\%, that is, the valid data with low PSNR gain rate accounts for a large proportion. So we expect the network to focus more on data that has low PSNR gain rate. To this end, we propose a focal MSE loss function to fine tune the proposed network, which can be calculated by:
\begin{equation}
        L = \alpha_{q}(1 - R)^\gamma L_{rec} = \alpha_{q} \frac{L^{1+\gamma}_{rec}}{L_{init}} \\
\end{equation}
where $\alpha_q$ is a weighting factor over QP value, and $\gamma$ is focusing parameter. Herein $\alpha_q = 0.1, 0.25, 0.3, 0.35$ for four QPs respectively and $\gamma=1$. Table \ref{focal} shows the coding performances of three networks over test sequences. QPALF-\uppercase\expandafter{\romannumeral1}, QPALF-\uppercase\expandafter{\romannumeral2}, QPALF-\uppercase\expandafter{\romannumeral3} denote QPALF without fine tuning, QPALF fine tuned by MSE and QPALF fine tuned by focal MSE respectively. As we can see, QPALF-\uppercase\expandafter{\romannumeral3} achieves more bit-rate saving than QPALF-\uppercase\expandafter{\romannumeral2}, which demonstrates the effectiveness of focal MSE.
\begin{table}[]
    \centering
    \caption{The coding performance of three QPALF networks}
% Table generated by Excel2LaTeX from sheet 'RES'
\begin{tabular}{cccc}
\hline
\multicolumn{ 1}{c}{\multirow{2}{*}{\bf Class}} &     \multicolumn{ 3}{c}{BD-Rate(\%)} \\
\cmidrule{2-4}
\multicolumn{ 1}{c}{{\bf }} & {\bf QPALF-\uppercase\expandafter{\romannumeral1}} & {\bf QPALF-\uppercase\expandafter{\romannumeral2}} & {\bf QPALF-\uppercase\expandafter{\romannumeral3}} \\
\hline
        A1 &      -1.54 &      -1.73 &      -2.02 \\

        A2 &      -1.98 &      -2.18 &      -2.29 \\

         B &      -3.14 &      -3.30 &      -3.32 \\

         C &      -4.49 &      -4.61 &      -4.73 \\

         D &      -5.48 &      -5.60 &      -5.70 \\

         E &      -5.43 &      -5.66 &      -5.84 \\
\hline
   Average &      \bf -3.75 &      \bf -3.91 &     \bf -4.03 \\
\hline
\end{tabular}  
    \label{focal}
\end{table}

\textbf{Training Detail}. The widely adopted deep learning framework Pytorch \cite{Pytorch} is utilized to train our models. We use Adam \cite{Adam} optimization to train these models and a batch size of 64. The learning rate discounts 0.5 every 25 epochs. The training takes 100 epochs in total. For QP-separate models, we first train the model for QP=37 and then use it to initialize the parameters of the networks with smaller QP. The initial learning rate is $10^{-4}$ for QP=37 and $10^{-5}$ for other QPs. For QP-combined models, the initial learning rate is $10^{-4}$ and the fine-tune process takes 50 epochs with learning rate $10^{-5}$. All models are trained on NVIDIA Titan X (Pascal) GPUs. 

\textbf{Implementation}. We integrate the QPALF into VVC as an additional tool of in-loop filters between DBF and SAO. To get better performance for video coding, a frame level flag is signaled in the bitstream to indicate whether QPALF is enabled for this frame in the decoder. When the reduction of RD cost is greater than 0, the flag will be enabled and the QPALF will be applied for the frame on luma component.
\begin{table}[]
    \centering
    \caption{The BD-Rate of different models on Y channel under AI configuration}
% Table generated by Excel2LaTeX from sheet 'RES'
\begin{tabular}{cccccccc}
\hline
\multicolumn{ 1}{c}{\multirow{2}{*}{\bf Class}} & \multicolumn{ 2}{c}{\multirow{2}{*}{\bf Sequence}} &                               \multicolumn{ 5}{c}{BD-Rate(\%)} \\
\cmidrule{4-8}
\multicolumn{ 1}{c}{{\bf }} & \multicolumn{ 2}{c}{{\bf }} &      RHCNN &        DRNLF  &      QPMLF &    QPALF-S &      QPALF \\
\hline
\multicolumn{ 1}{c}{\multirow{3}{*}{A1}} & \multicolumn{ 2}{c}{Tango2} &      -0.62 &      -0.63 &      -0.78 &      -0.62 &      -1.86 \\

\multicolumn{ 1}{c}{} & \multicolumn{ 2}{c}{Campfire} &      -0.82 &      -1.32 &      -0.79 &      -1.42 &      -2.01 \\

\multicolumn{ 1}{c}{} & \multicolumn{ 2}{c}{FoodMarket4} &      -0.74 &      -0.09 &      -0.29 &      -0.89 &      -2.20 \\
\hline
\multicolumn{ 1}{c}{\multirow{3}{*}{A2}} & \multicolumn{ 2}{c}{CatRobot} &      -1.10 &      -2.20 &      -1.97 &      -2.28 &      -3.39 \\

\multicolumn{ 1}{c}{} & \multicolumn{ 2}{c}{DaylightRoad2} &      -0.43 &       0.07 &      -0.17 &       1.02 &      -0.50 \\

\multicolumn{ 1}{c}{} & \multicolumn{ 2}{c}{ParkRunning3} &      -1.01 &      -1.96 &      -1.56 &      -2.04 &      -2.99 \\
\hline
\multicolumn{ 1}{c}{\multirow{5}{*}{B}} & \multicolumn{ 2}{c}{RitualDance} &      -1.88 &      -4.29 &      -4.03 &      -4.85 &      -6.32 \\

\multicolumn{ 1}{c}{} & \multicolumn{ 2}{c}{MarketPlace} &      -1.34 &      -2.33 &      -2.09 &      -2.64 &      -3.58 \\

\multicolumn{ 1}{c}{} & \multicolumn{ 2}{c}{BasketballDrive} &      -0.73 &      -1.63 &      -1.05 &      -1.82 &      -2.84 \\

\multicolumn{ 1}{c}{} & \multicolumn{ 2}{c}{BQTerrace} &      -0.64 &      -1.38 &      -1.02 &      -1.56 &      -2.06 \\

\multicolumn{ 1}{c}{} & \multicolumn{ 2}{c}{Cactus} &      -0.84 &      -2.07 &      -1.71 &      -1.69 &      -1.78 \\
\hline
\multicolumn{ 1}{c}{\multirow{4}{*}{C}} & \multicolumn{ 2}{c}{BasketballDrill} &      -2.29 &      -5.43 &      -4.50 &      -5.76 &      -7.48 \\

\multicolumn{ 1}{c}{} & \multicolumn{ 2}{c}{BQMall} &      -1.93 &      -4.31 &      -3.73 &      -4.58 &      -5.49 \\

\multicolumn{ 1}{c}{} & \multicolumn{ 2}{c}{PartyScene} &      -1.22 &      -3.01 &      -2.57 &      -3.19 &      -3.62 \\

\multicolumn{ 1}{c}{} & \multicolumn{ 2}{c}{RaceHorsesC} &      -0.81 &      -1.75 &      -1.39 &      -1.81 &      -2.31 \\
\hline
\multicolumn{ 1}{c}{\multirow{4}{*}{D}}& \multicolumn{ 2}{c}{BasketballPass} &      -2.10 &      -5.24 &      -4.42 &      -5.67 &      -6.76 \\

\multicolumn{ 1}{c}{} & \multicolumn{ 2}{c}{BlowingBubbles} &      -1.59 &      -3.64 &      -3.19 &      -3.87 &      -4.45 \\

\multicolumn{ 1}{c}{} & \multicolumn{ 2}{c}{BQSquare} &      -1.93 &      -5.12 &      -4.40 &      -5.28 &      -6.20 \\

\multicolumn{ 1}{c}{} & \multicolumn{ 2}{c}{RaceHorses} &      -2.03 &      -4.54 &      -4.21 &      -4.67 &      -5.40 \\
\hline
\multicolumn{ 1}{c}{\multirow{3}{*}{E}} & \multicolumn{ 2}{c}{FourPeople} &      -2.05 &      -4.73 &      -4.08 &      -5.09 &      -6.49 \\

\multicolumn{ 1}{c}{} & \multicolumn{ 2}{c}{Johnny} &      -1.63 &      -3.90 &      -3.17 &      -4.12 &      -5.72 \\

\multicolumn{ 1}{c}{} & \multicolumn{ 2}{c}{KristenAndSara} &      -1.59 &      -3.95 &      -3.31 &      -4.24 &      -5.31 \\
\hline
\multicolumn{ 3}{c}{{\bf Average All}} & {\bf -1.54} & {\bf -2.88} & {\bf -2.47} & {\bf -3.05} & {\bf -4.03} \\
\hline
\end{tabular}  
    \label{RD performance}
\end{table}
\Section{3. Experiment} \label{experiment}
\subsection*{Experimental Setting}
In our experiments, all approaches for in-loop filtering are incorporated into the VVC reference software VTM6.0. The Libtorch \cite{Pytorch} library is integrated into VTM 6.0 to perform the in-loop filtering with the different models. Four typical QP values are tested, including 22, 27, 32, 37. We use the AI configuration suggested by VVC common test condition (CTC) \cite{CTC}.
The anchor for all experiments is VTM6.0 with all built-in in-loop filters enabled. The coding efficiency is evaluated on standard video sequences from class A1 to class E recommended by JVET. The BD-Rate \cite{BD-Rate} are referred to measure the coding performance. We only train and apply the models on Y channel, but our approach can be extended to any arbitrary number of channels.
\subsection*{Evaluation on VVC Test Sequences}
\begin{figure}[t]
\centering
% \subfigure[FourPeople]{
% \begin{minipage}[t]{0.5\linewidth}
% \centering
% \includegraphics[width=1.75in]{img/FourPeople_multi_qp_nomap-eps-converted-to.pdf}
% % \caption*{VTM(28.15, 0.8643)}
% \end{minipage}%
% }%
\subfigure[FourPeople]{
\begin{minipage}[t]{0.4\linewidth}
\centering
\includegraphics[width=\linewidth]{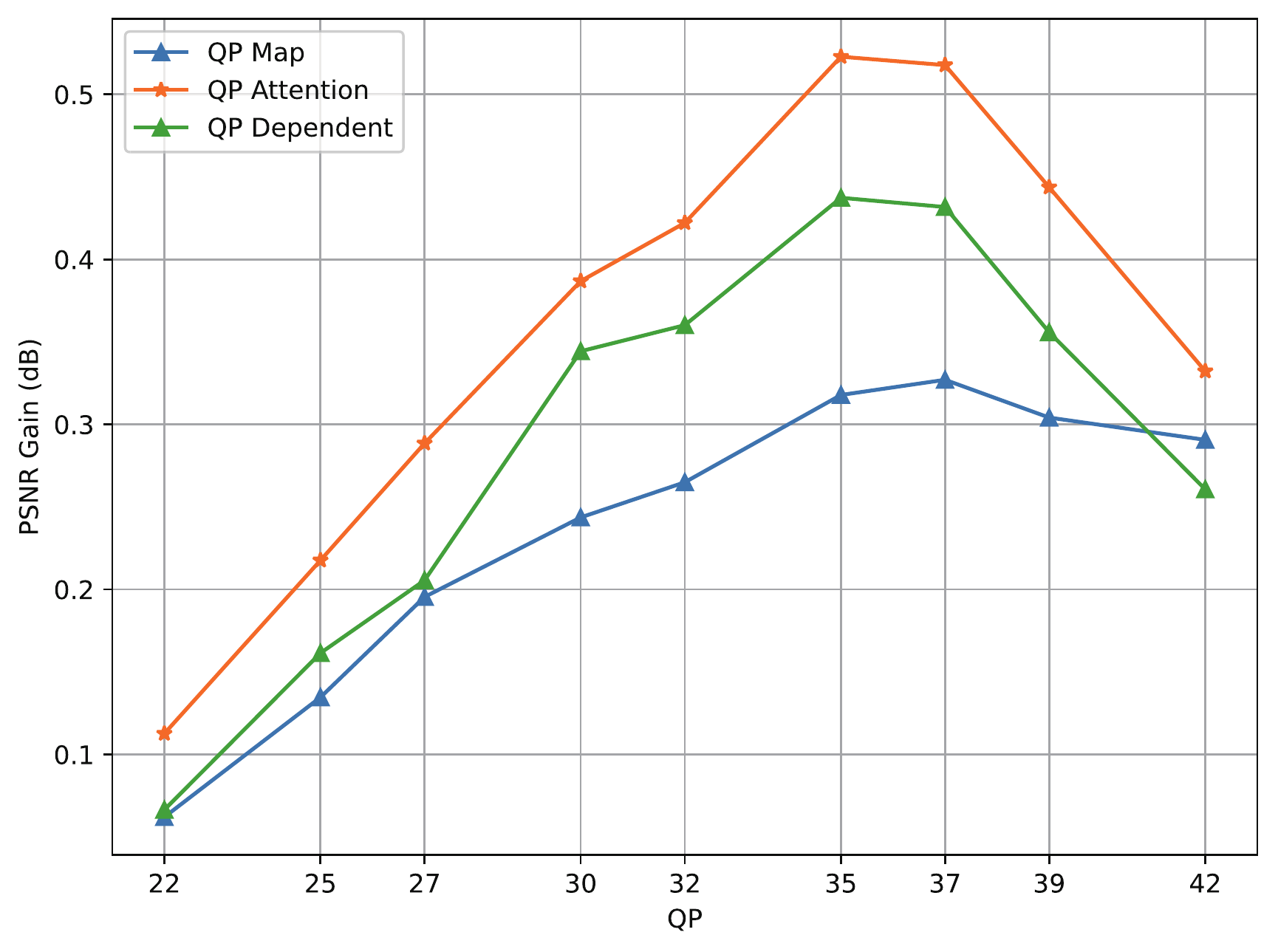}
% \caption*{RHCNN(28.24, 0.8666)}
\end{minipage}
}%
\subfigure[BQMall]{
\begin{minipage}[t]{0.4\linewidth}
\centering
\includegraphics[width=\linewidth]{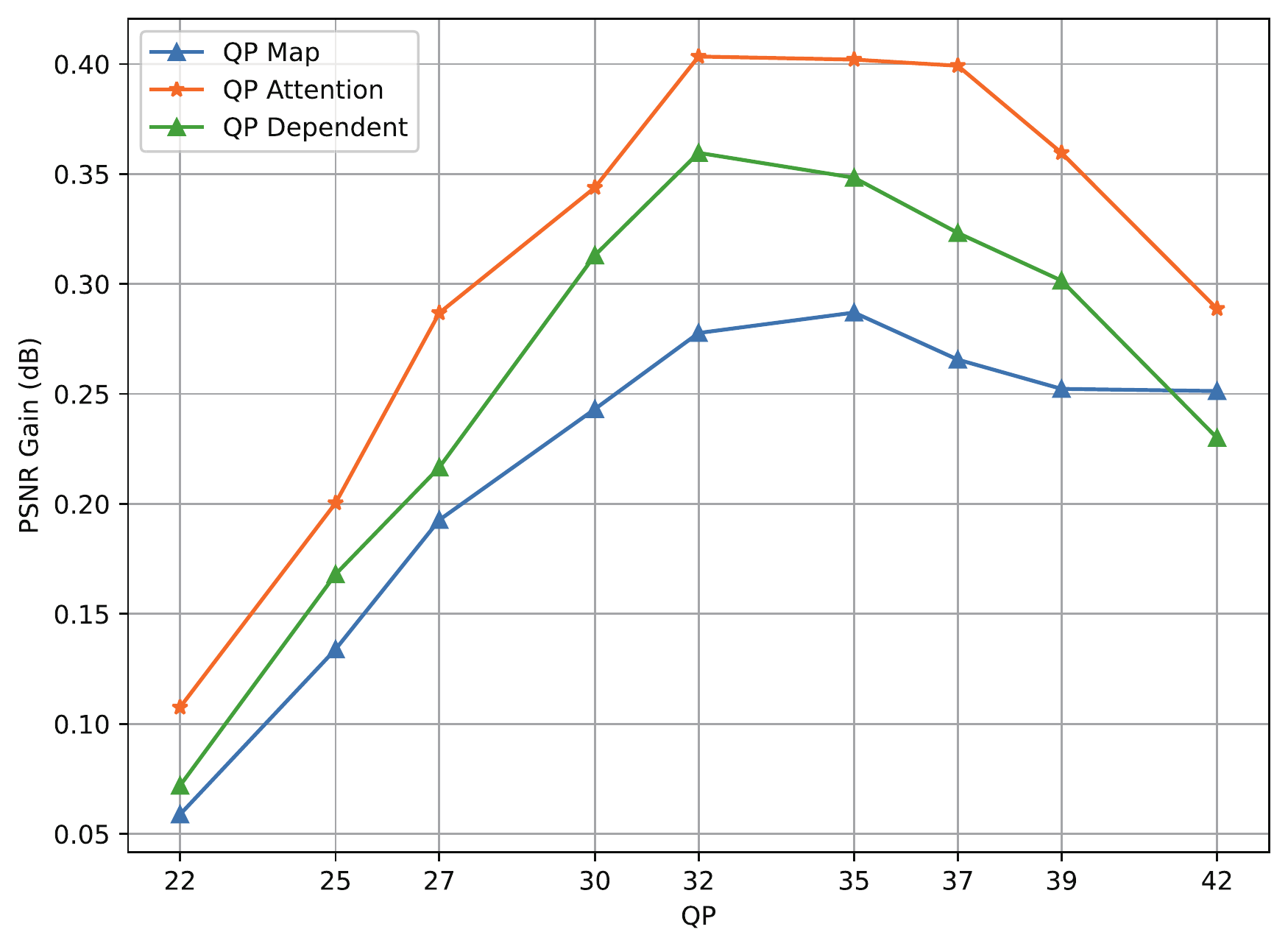}
% \caption*{Ours(\textbf{28.42, 0.8714})}
\end{minipage}
}%
% \subfigure[BQMall]{
% \begin{minipage}[t]{0.5\linewidth}
% \centering
% \includegraphics[width=1.75in]{img/BQMall_multi_qp_nomap-eps-converted-to.pdf}
% % \caption*{DRN(28.32, 0.8682)}
% \end{minipage}
% }%
\centering
\caption{PSNR gain of three methods on multiple QPs. (a) FourPeople; (b) BQMall. }
\label{curve}
\end{figure}

\textbf{RD performance}. First, we compare our QPALF with VTM baseline, and two CNN based in-loop filters, RHCNN \cite{RHCNN} and DRNLF \cite{DRNLF}. For a fair comparison, we also train the models using our dataset and integrate the trained model into VTM6.0 between DBF and SAO. The results are displayed in Table \ref{RD performance}. It can be seen obviously, our QPALF further improves the coding efficiency, which obtains 4.03\% bit-rate saving overall for the luma component on all the test sequences. To further verify the efficiency of the proposed model, we also compare our QPALF with QPMLF and QPALF-D. QPALF-D is the QPALF model trained separately on four QPs. QPMLF is the QPALF without QPAM and use the QP map method \cite{CNNLF}. As we can see, compared with QP-separate model QPALF-S, the performance of the QP-combined model QPMLF degrades while our model QPALF achieves even better coding performance. Moreover, the PSNR gain of three models on multiple QPs is also depicted in Figure \ref{curve}. We can observe that our QPAM obtains the highest PSNR gain over all QPs, which also demonstrates the generalization ability and robustness of the proposed method. (Since the models are trained on only four QPs, \{22, 27, 32, 37\}, we first map other QP values to the four QPs.)

\begin{figure*}[thbp]
\centering
\subfigure{
\begin{minipage}[t]{0.2\linewidth}
\centering
\includegraphics[width=\linewidth]{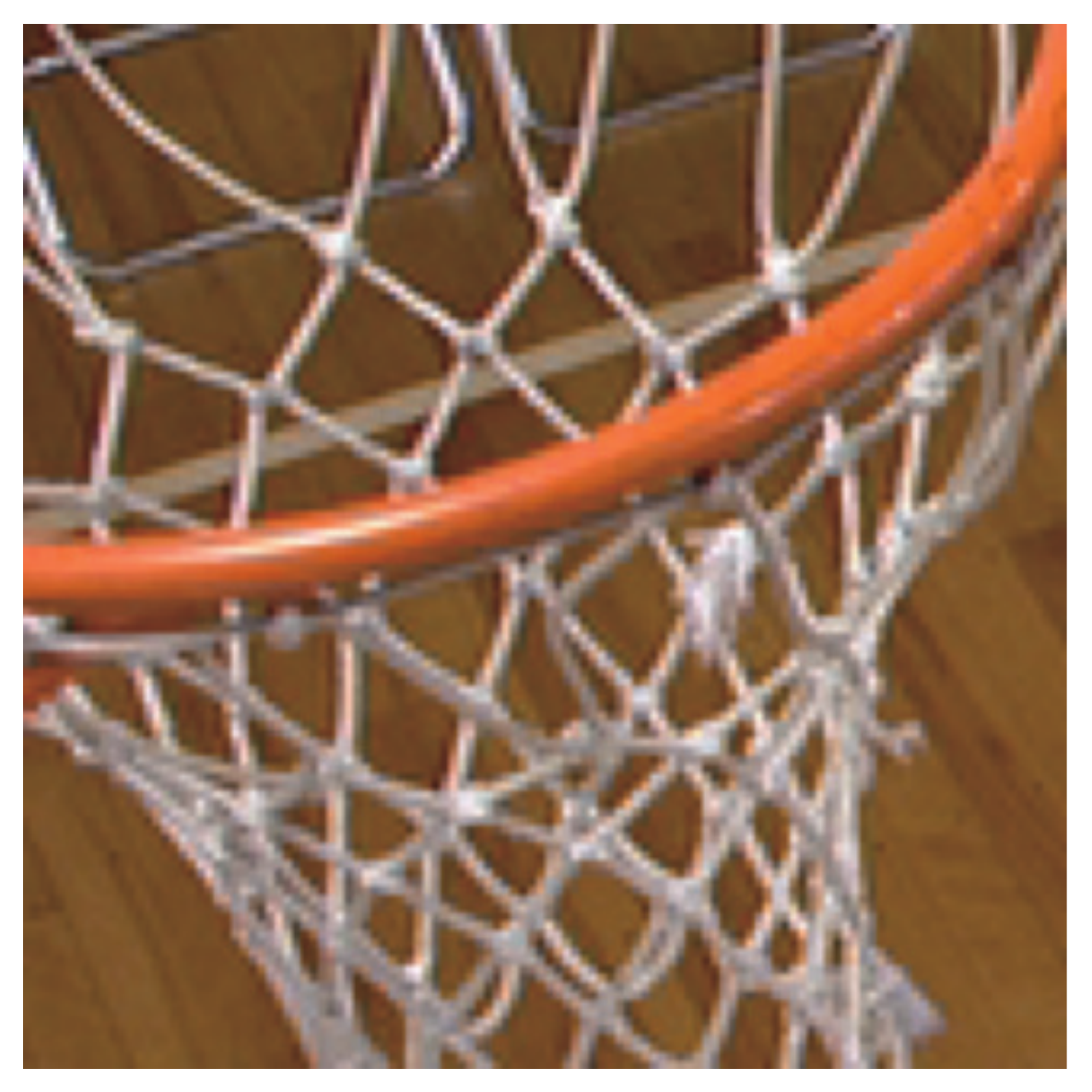}
\caption*{\centerline{Ground Truth} \\ \centerline{(PSNR,SSIM)}}
\end{minipage}%
}%
\subfigure{
\begin{minipage}[t]{0.2\linewidth}
\centering
\includegraphics[width=\linewidth]{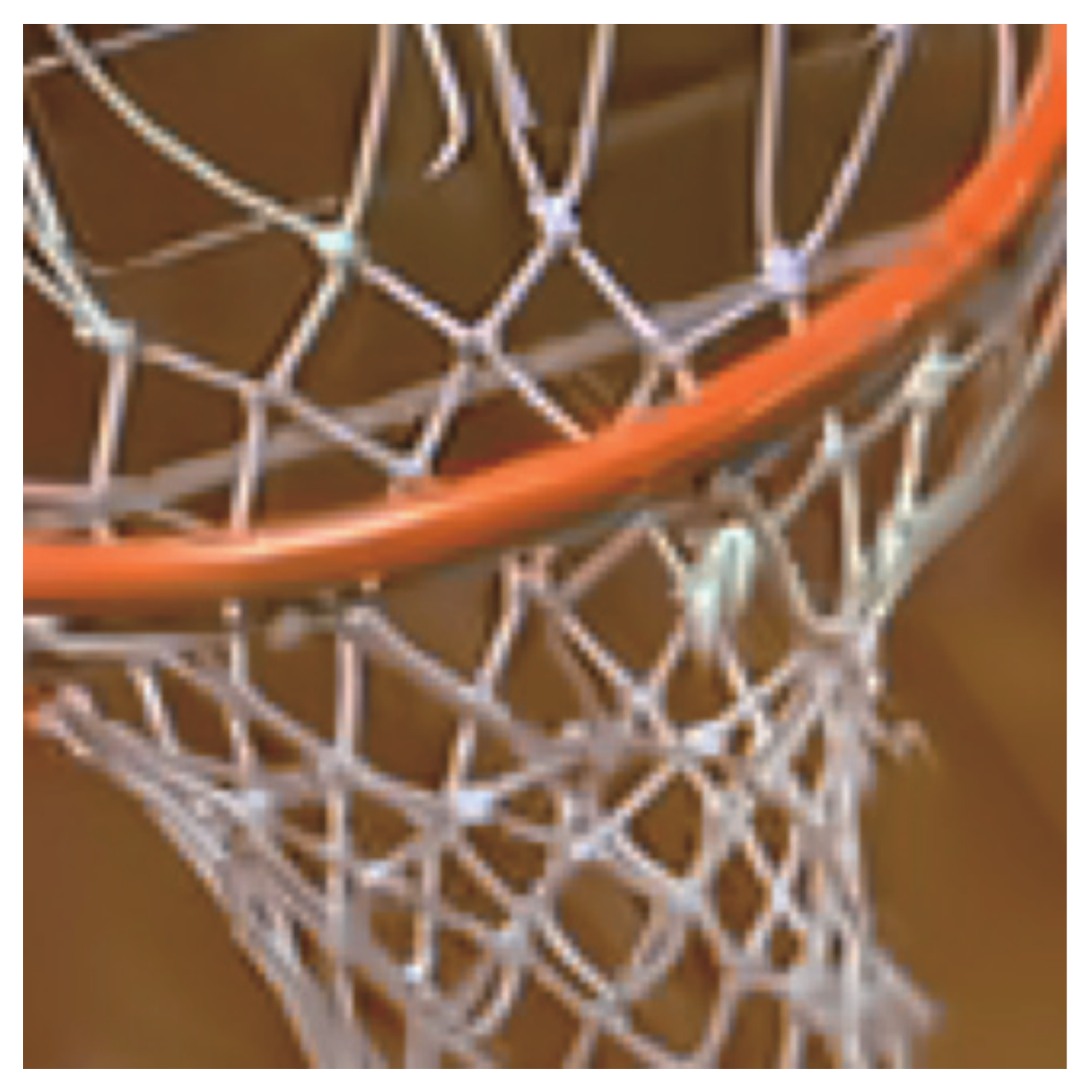}
\caption*{\centerline{VTM} \\ \centerline{(26.16, 0.9045)}}
\end{minipage}%
}%
\subfigure{
\begin{minipage}[t]{0.2\linewidth}
\centering
\includegraphics[width=\linewidth]{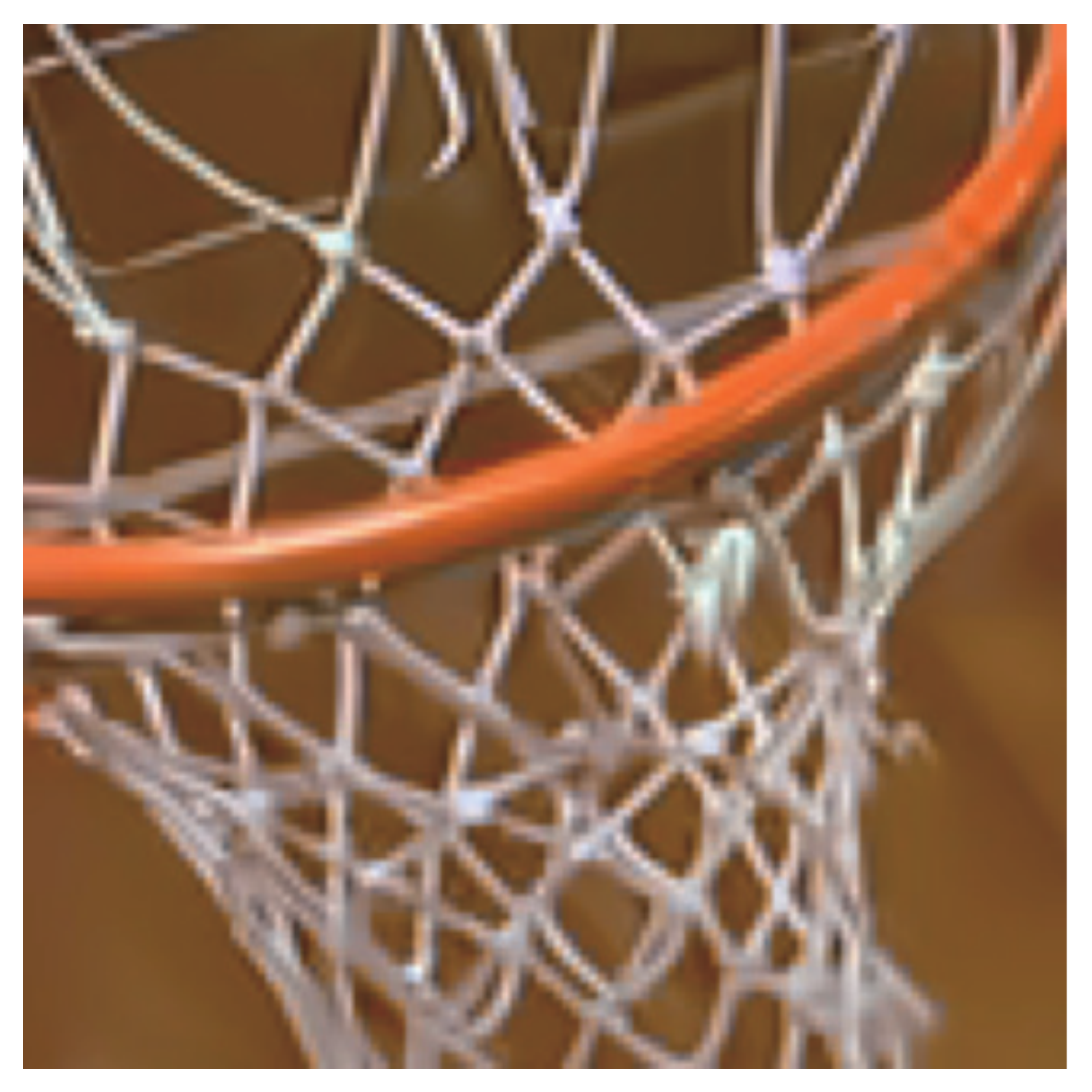}
\caption*{\centerline{RHCNN} \\ \centerline{(26.29, 0.9073)}}
\end{minipage}
}%
\subfigure{
\begin{minipage}[t]{0.2\linewidth}
\centering
\includegraphics[width=\linewidth]{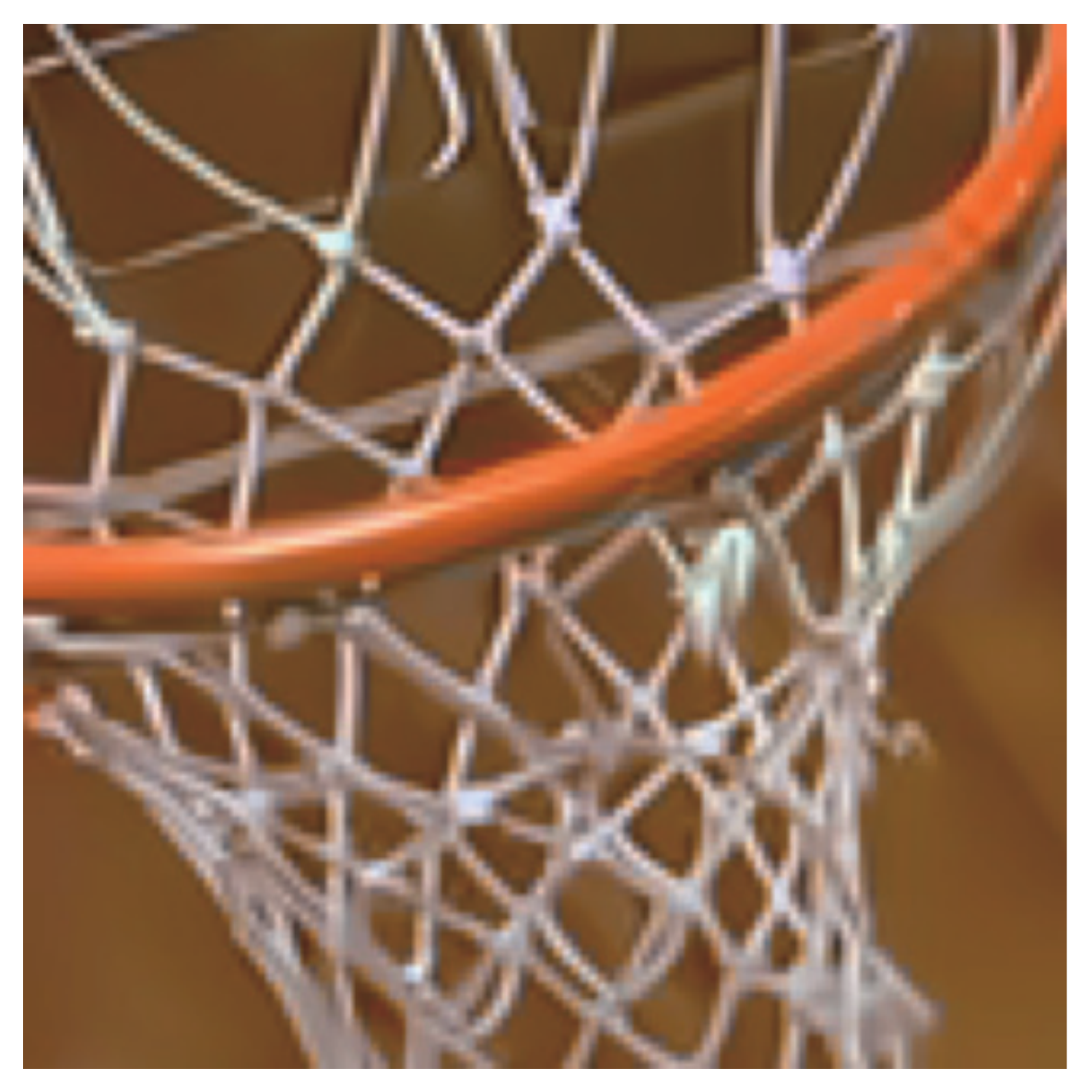}
\caption*{\centerline{DRNLF} \\ \centerline{(26.56, 0.9104)}}
\end{minipage}
}%
\subfigure{
\begin{minipage}[t]{0.2\linewidth}
\centering
\includegraphics[width=\linewidth]{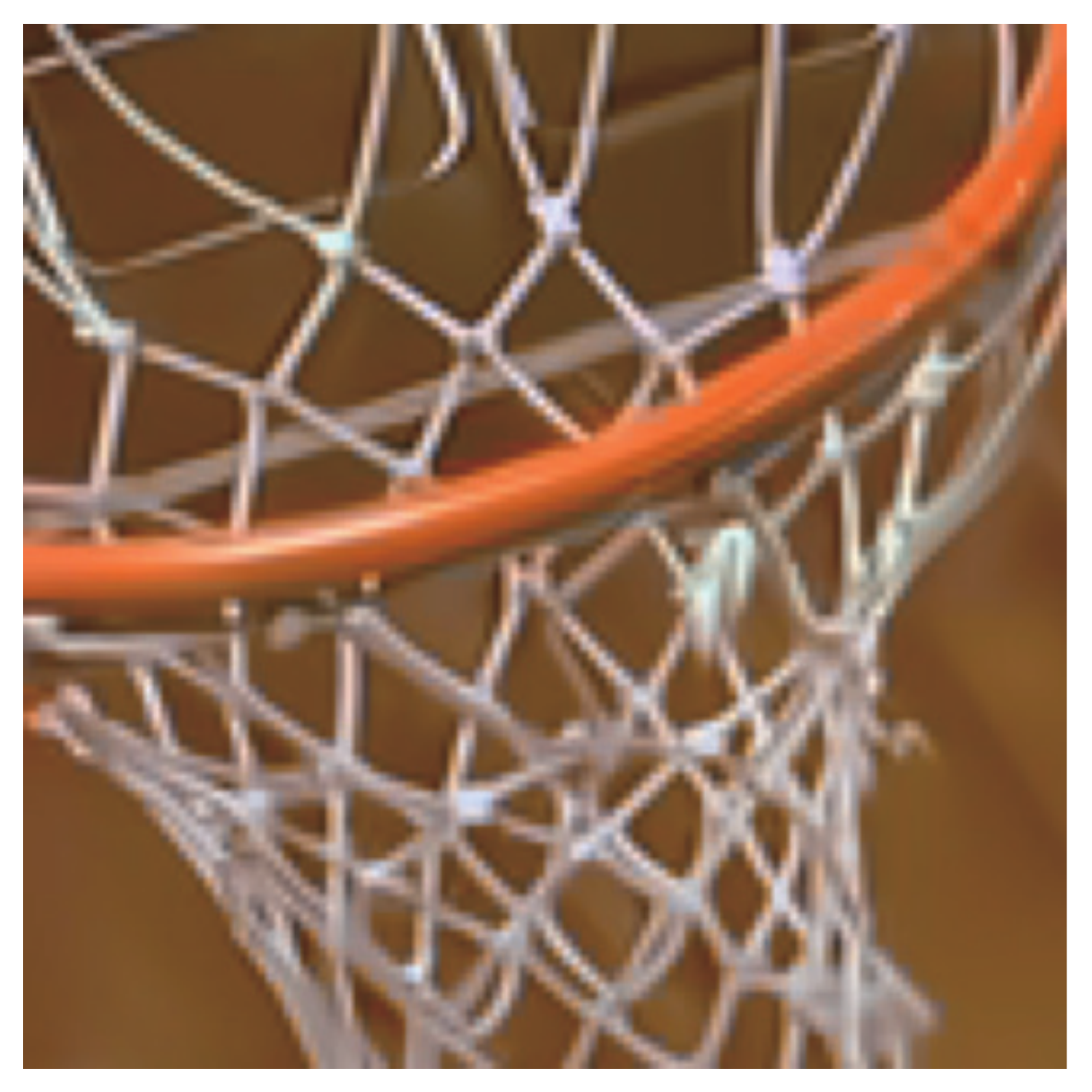}
\caption*{\centerline{Ours}\\\centerline{(\textbf{26.66, 0.9120})}}
\end{minipage}
}%

\subfigure{
\begin{minipage}[t]{0.2\linewidth}
\centering
\includegraphics[width=1.2in]{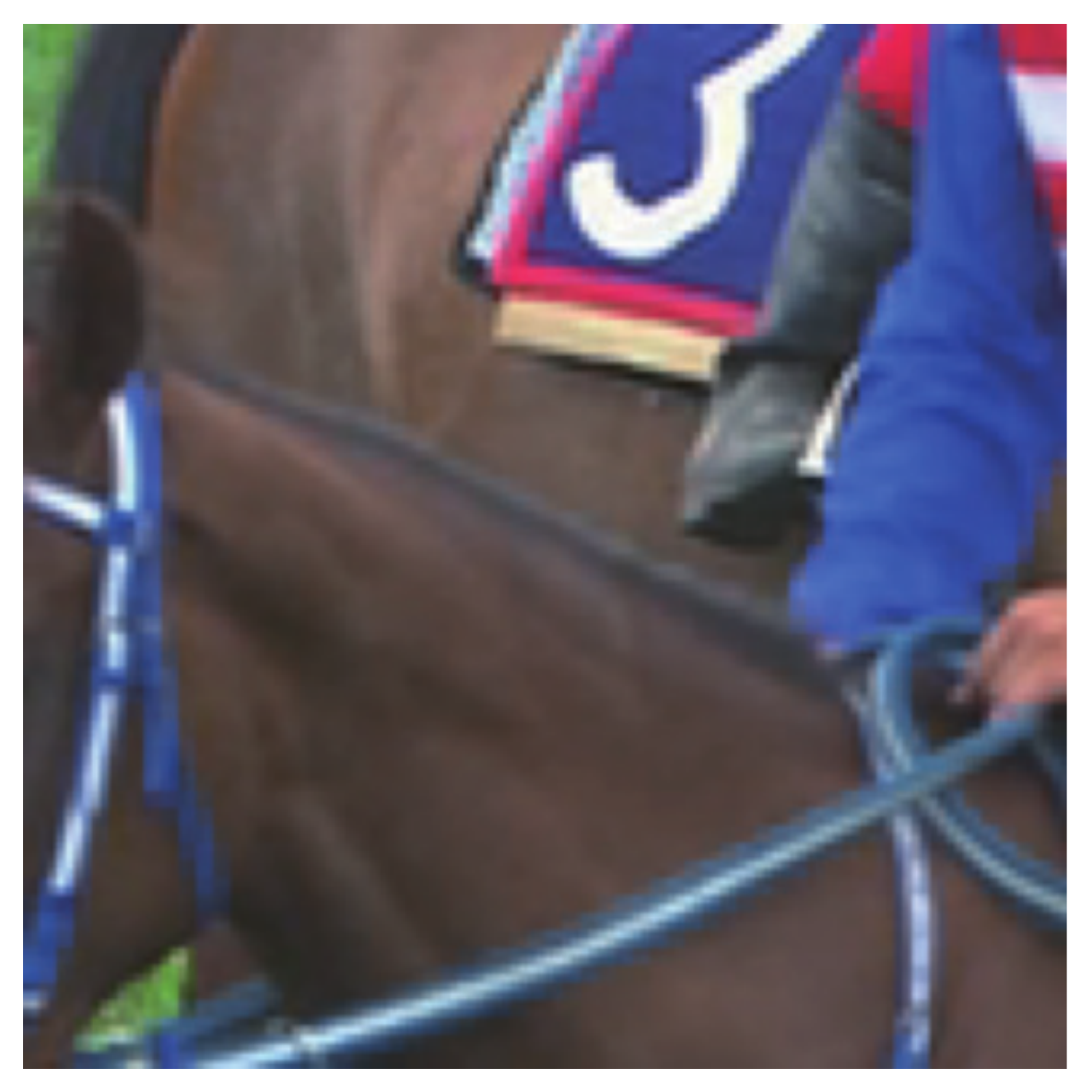}
\caption*{\centerline{Ground Truth} \\ \centerline{(PSNR,SSIM)}}
\end{minipage}%
}%
\subfigure{
\begin{minipage}[t]{0.2\linewidth}
\centering
\includegraphics[width=1.2in]{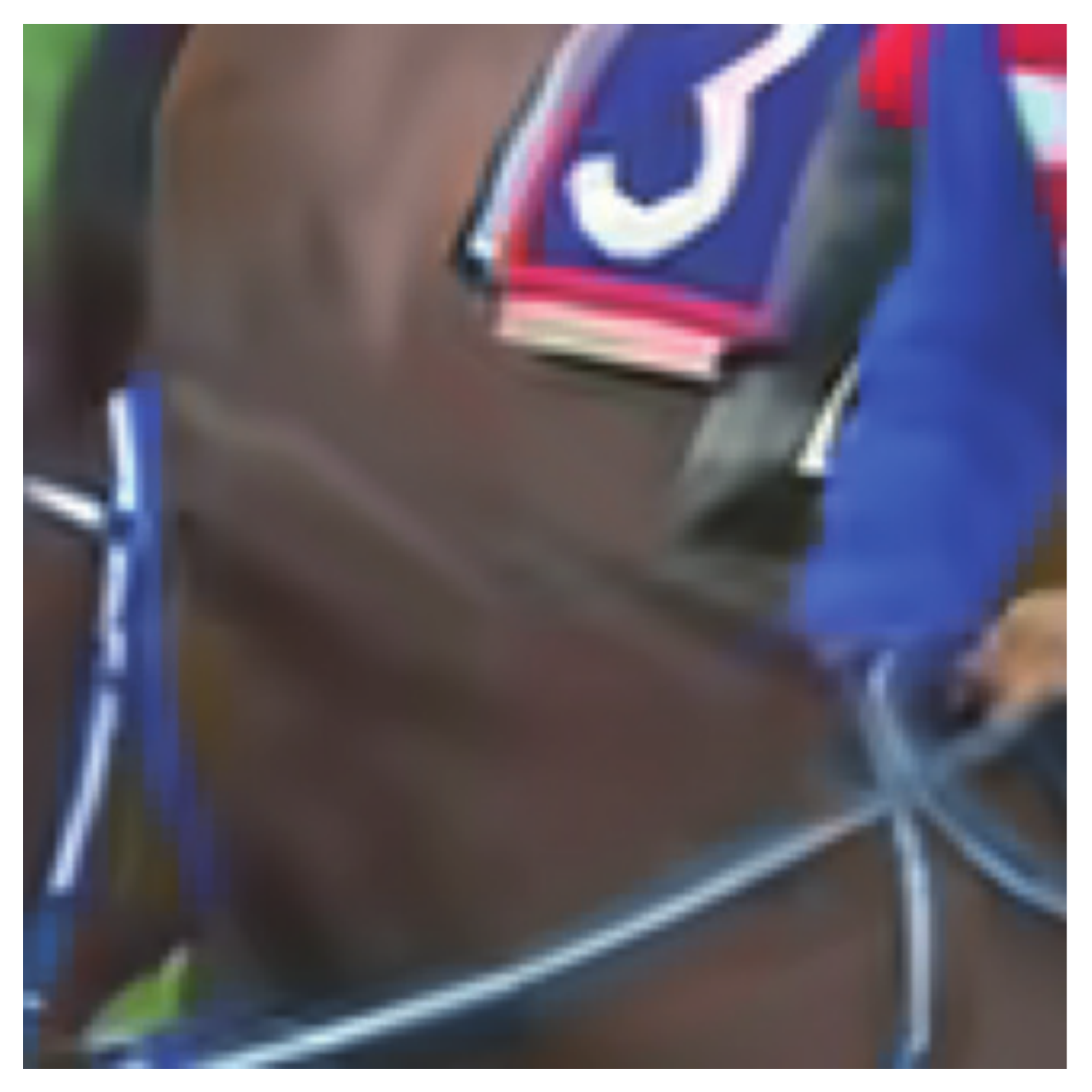}
\caption*{\centerline{VTM} \\ \centerline{(28.15, 0.8643)}}
\end{minipage}%
}%
\subfigure{
\begin{minipage}[t]{0.2\linewidth}
\centering
\includegraphics[width=1.2in]{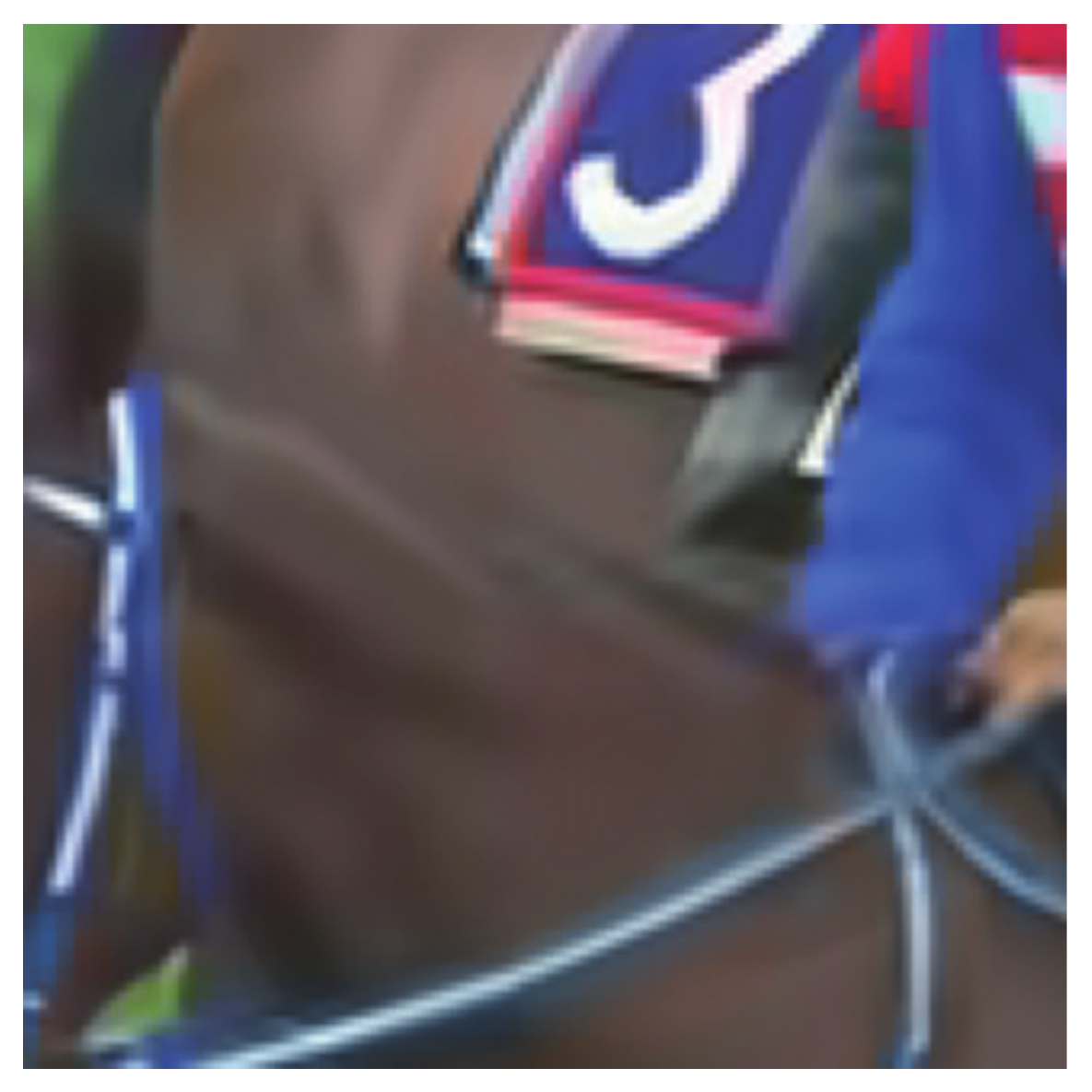}
\caption*{\centerline{RHCNN} \\ \centerline{(28.24, 0.8666)}}
\end{minipage}
}%
\subfigure{
\begin{minipage}[t]{0.2\linewidth}
\centering
\includegraphics[width=1.2in]{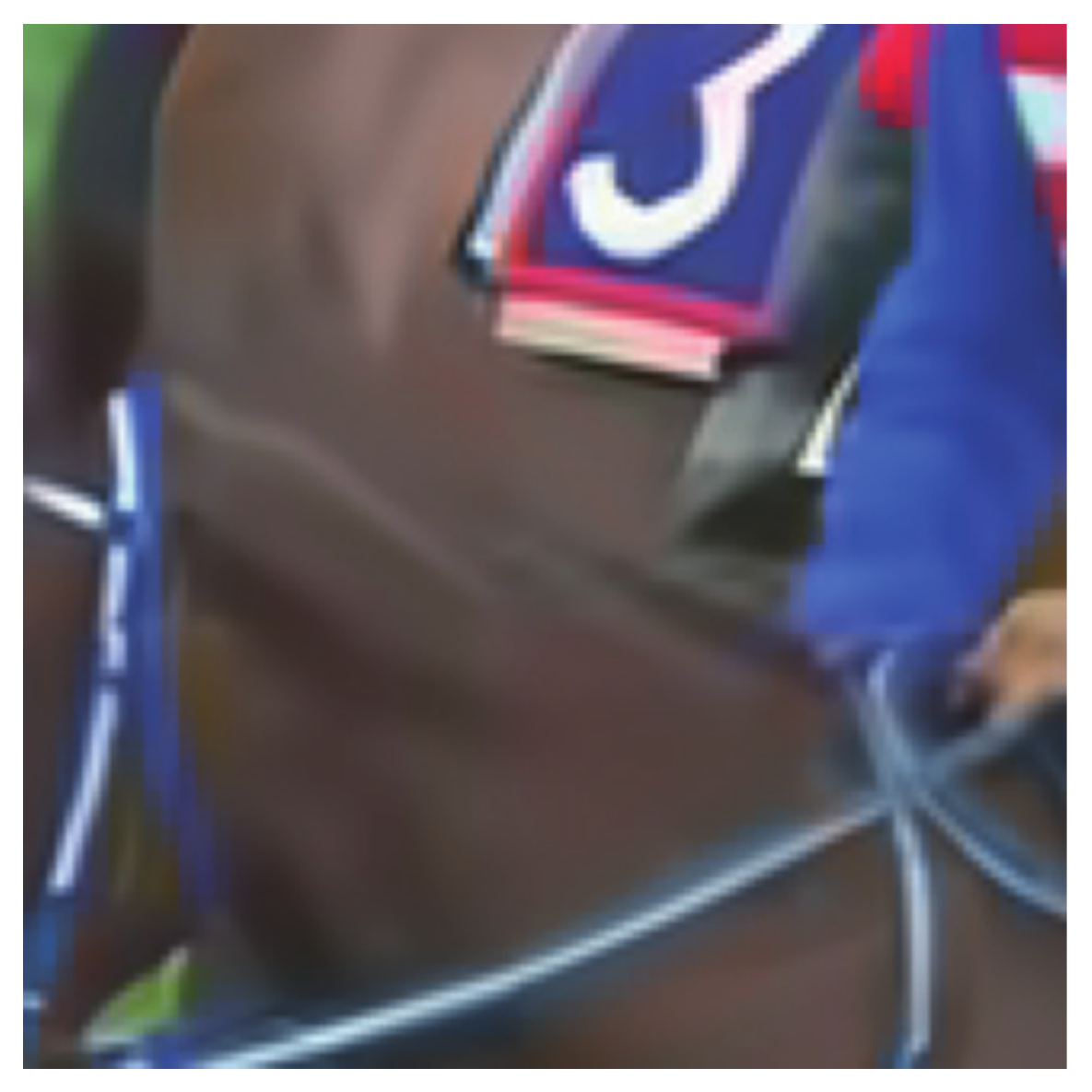}
\caption*{\centerline{DRNLF} \\ \centerline{(28.32, 0.8682)}}
\end{minipage}
}%
\subfigure{
\begin{minipage}[t]{0.2\linewidth}
\centering
\includegraphics[width=1.2in]{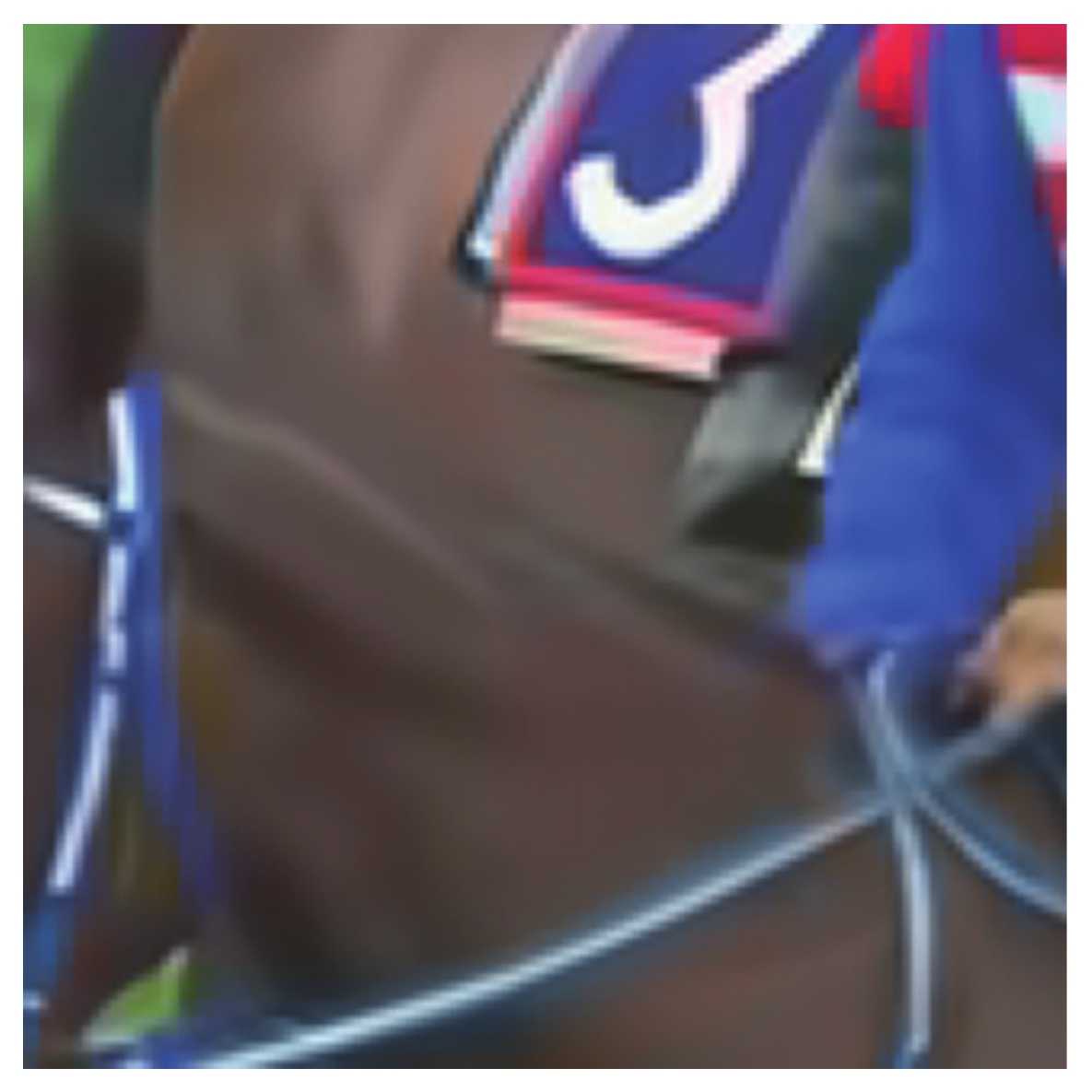}
\caption*{\centerline{Ours} \\ \centerline{(\textbf{28.42, 0.8714})}}
\end{minipage}
}%
\centering
\caption{Subjective image quality comparison on sequences BasketballDrill (Class D) and RaceHorses (Class C) at QP = 37}
\label{Sub1}
\end{figure*}

\begin{figure*}[thbp]
\centering
\subfigure[VTM]{
\begin{minipage}[t]{0.25\linewidth}
\centering
\includegraphics[width=1.6in]{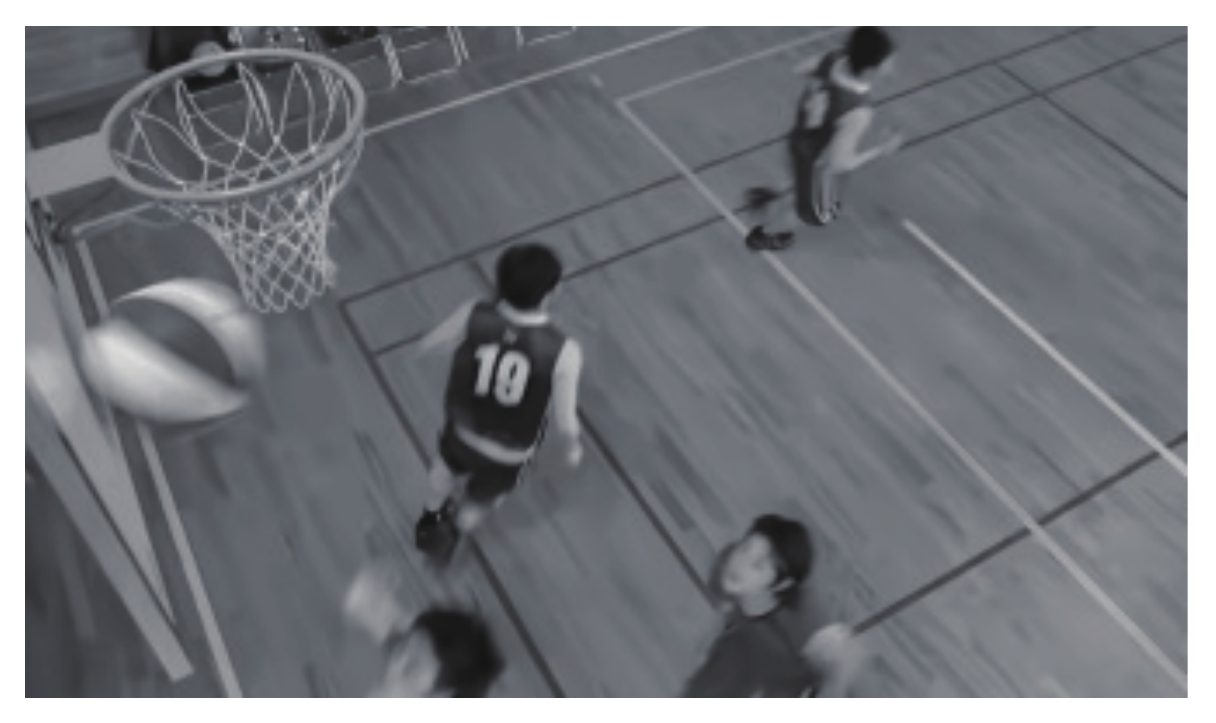}
% \caption*{VTM(28.15, 0.8643)}
\end{minipage}%
}%
\subfigure[RHCNN]{
\begin{minipage}[t]{0.25\linewidth}
\centering
\includegraphics[width=1.6in]{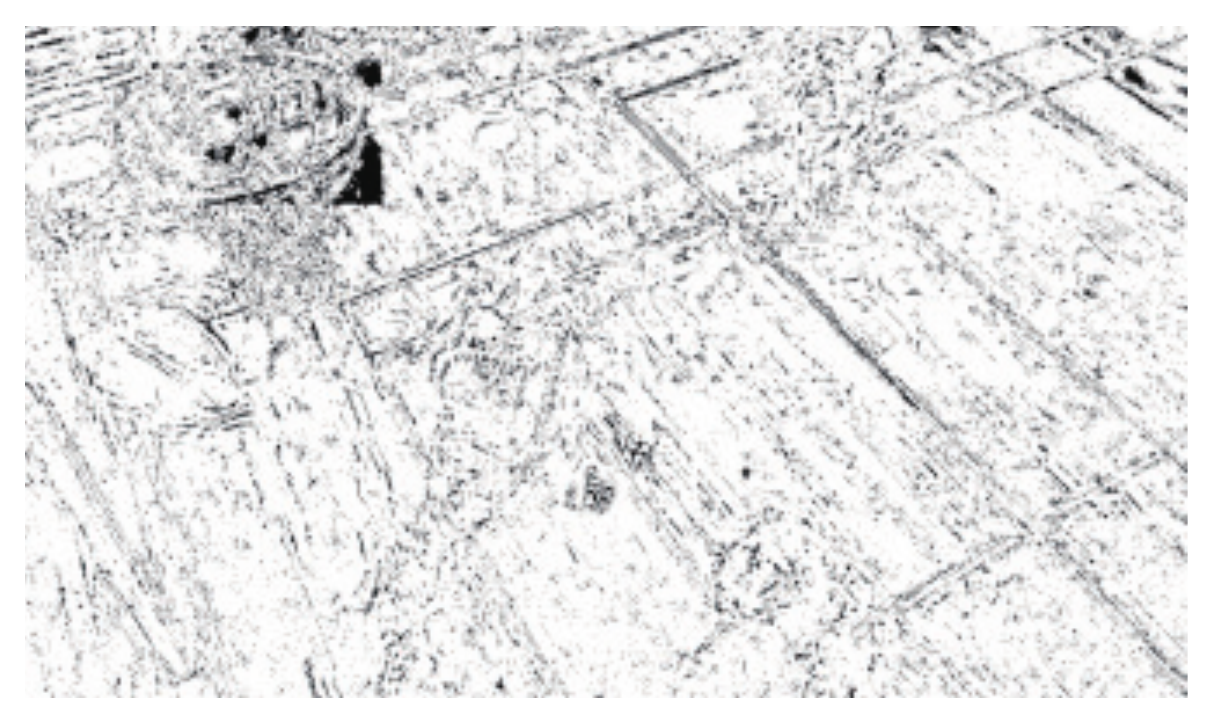}
% \caption*{RHCNN(28.24, 0.8666)}
\end{minipage}
}%
\subfigure[DRNLF]{
\begin{minipage}[t]{0.25\linewidth}
\centering
\includegraphics[width=1.6in]{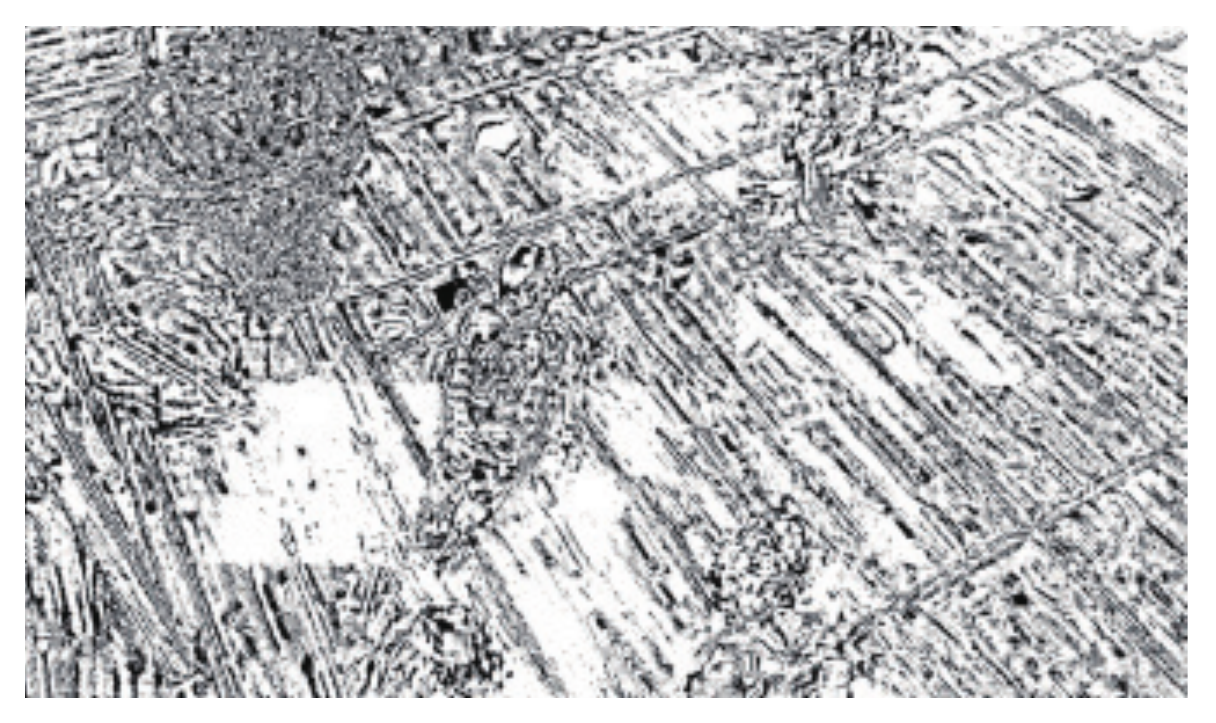}
% \caption*{DRN(28.32, 0.8682)}
\end{minipage}
}%
\subfigure[Ours]{
\begin{minipage}[t]{0.25\linewidth}
\centering
\includegraphics[width=1.6in]{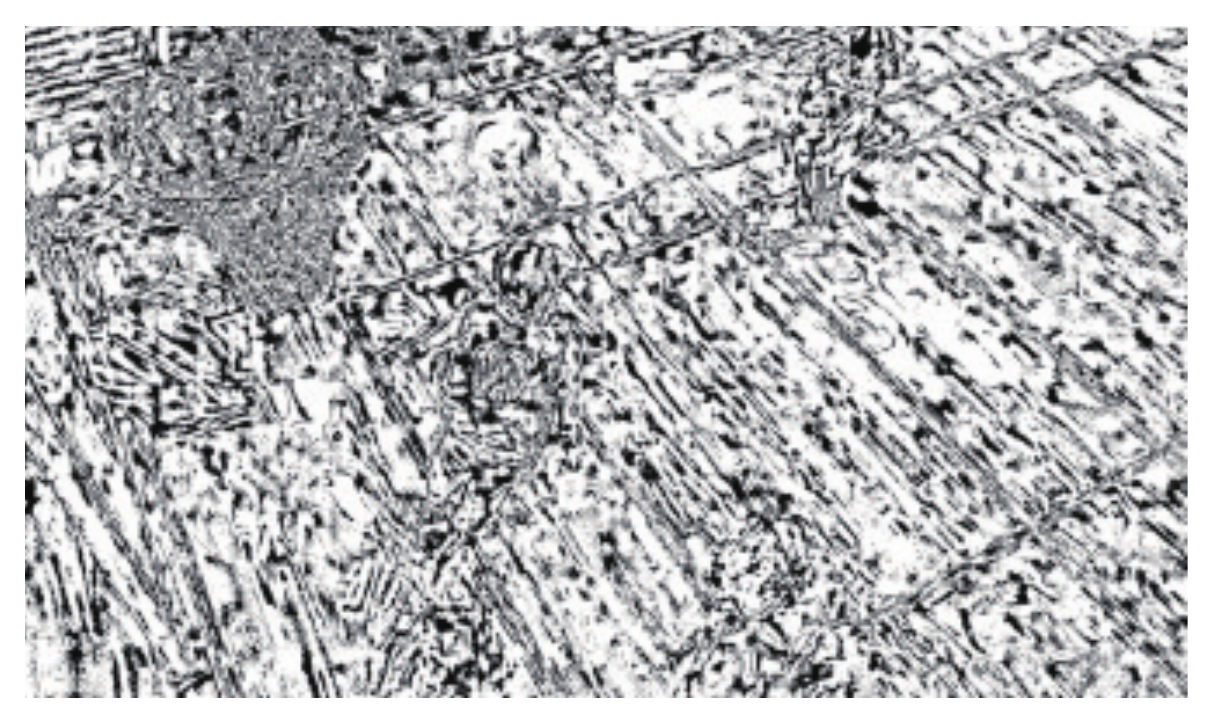}
% \caption*{Ours(\textbf{28.42, 0.8714})}
\end{minipage}
}%
\centering
\caption{Residual map of different methods over VTM baseline.}
\label{Sub2}
\end{figure*}

\textbf{Subjective evaluation}. Figure \ref{Sub1} illustrates the subjective visual quality comparison among all four approaches. It can be observed that the images enhanced by our approach remain less distortion than those by other approaches, e.g., the clearer edge of the basketball net line. In Figure \ref{Sub2}, we display the residual map of three methods over VTM baseline. Compared with RHCNN and DRNLF, our method can restore more image texture details.  

\textbf{Complexity}. Table \ref{complexity} shows the average encode/decode complexity increase and parameters of different models on Intel(R) Xeon(R) CPU E5-2697v4 and Titan X (Pascal). All of the neural networks are  conducted  with  GPU  acceleration. The complexity increase is calculated by $\Delta T = (\hat{T} - T) / T$, where $\hat{T}$ is the encode/decode time with integrating the models, $T$ is the original encode/decode time. Our proposed QPALF has less model parameters overall. Compared with QPMLF, our model achieves much better RD performance with little complexity increase.


\begin{table}[hbtp]
    \centering
    \caption{Average complexity increase and parameters of different models}
\begin{tabular}{ccccc}
\hline
    Method &  {\bf $\Delta$ET} &  {\bf $\Delta$DT} & \multicolumn{ 2}{c}{\bf \#Params } \\
\hline
     RHCNN &     5.43\% &    10695.9\% & \multicolumn{ 2}{c}{ $6.79M\times4$ } \\

      DRN &     3.29\% &     7808.0\% & \multicolumn{ 2}{c}{$671.30k \times4$} \\
     QPMLF &     4.41\%  & 8232.2\% & \multicolumn{ 2}{c}{$838.98k \times1$}\\
      QPALF &     4.70\% &    8428.6\% & \multicolumn{ 2}{c}{$905.22k\times1$} \\
\hline
\end{tabular}  
    \label{complexity}
\end{table}

\Section{4. Conclusion and Future Work}\label{conclusion}
In this paper, an efficient QP variable CNN based in-loop filter for VVC is proposed. With the proposed QPAM, the QPALF can be adaptive to different QPs while achieving better RD performance. And a focal MSE is introduced to train a more robust model. Experimental results demonstrate that our QPALF can significantly improve the coding efficiency, which outperforms other CNN based methods. Moreover, the proposed QPAM has wide applicability and stronger scalability which can be easily to implement in networks and extend to other types. In our future work, we will extend our model to inter mode and speed up our QPALF.

\paragraph{Acknowledgment}
This work was supported by National Natural Foundation of China under contract No. 61671025

\Section{References}
\bibliographystyle{IEEEbib}
\bibliography{main}

\end{document}